\documentclass[sigconf]{acmart}

\usepackage{booktabs} 
\usepackage{booktabs} 
\usepackage{multirow}
\usepackage{booktabs}
\usepackage{subfigure}
 \usepackage{graphicx}
 \usepackage{algorithmic}
\usepackage[linesnumbered, ruled]{algorithm2e}
\usepackage{url}
\usepackage{multirow}
\usepackage{array}
\usepackage{amsfonts}
\usepackage{amsmath,bm}
 
\usepackage{pifont}
\newcommand{\cmark}{\ding{51}}%
\newcommand{\xmark}{\ding{55}}%
\usepackage{enumitem}

\usepackage{xcolor}
\usepackage{soul}
\usepackage{makecell}
\setul{}{0.5pt}
\usepackage{mathtools}
\SetKwInput{KwData}{Input}
\SetKwInput{KwResult}{Output}
\usepackage{amsthm}
\theoremstyle{definition}
\newtheorem{defn}{Definition}[section]

\acmDOI{10.475/123_4}

\acmISBN{123-4567-24-567/08/06}

\acmConference[XXX]{XXX}{XXX}{XXX} 
\acmYear{2018}
\copyrightyear{2018}

\begin{document}

\title{CARL: Content-Aware Representation Learning for Heterogeneous Networks}

\author{Chuxu Zhang}
\affiliation{%
  \institution{University of Notre Dame}
}
\email{czhang11@nd.edu}

\author{Ananthram Swami}
\affiliation{%
  \institution{Army Research Laboratory}
}
\email{ananthram.swami.civ@mail.mil}

\author{Nitesh V. Chawla}
\affiliation{%
  \institution{University of Notre Dame}
}
\email{nchawla@nd.edu}

\begin{abstract}
Heterogeneous networks not only present a challenge of heterogeneity in the types of nodes and relations, but also the attributes and content associated with the nodes. While recent works have looked at representation learning on homogeneous and heterogeneous networks, there is no work that has collectively addressed the following challenges:  (a) the heterogeneous structural information of the network consisting of multiple types of nodes and relations; (b) the unstructured semantic content (e.g., text) associated with nodes; and (c) online updates due to incoming new nodes in growing network. 
We address these challenges by developing a Content-Aware Representation Learning model (CARL). CARL performs joint optimization of heterogeneous SkipGram and deep semantic encoding for capturing both heterogeneous structural closeness and unstructured semantic relations among all nodes, as function of node content, that exist in the network. Furthermore, an additional online update module is proposed for efficiently learning representations of incoming nodes. Extensive experiments demonstrate that CARL outperforms state-of-the-art baselines in various heterogeneous network mining tasks, such as link prediction, document retrieval, node recommendation and relevance search. We also demonstrate the effectiveness of the CARL's online update module through a category visualization study. 
\end{abstract}


\maketitle

\section{Introduction}
Heterogeneous information networks (HetNets) \cite{sun2011pathsim,sun2012pathselclus}, e.g., academic networks, encode rich information through multi-typed nodes, relationships, and attributes or content associated with nodes. For example, the academic networks can represent human-human relationship (authors), human-object relationship (author-paper or author-venue or author-organization), and object-object relationship (paper-paper, paper-venue, paper-organization). The nodes in this case (human and object) can carry attributes or semantic content (such as paper abstract). Given the multi-typed nodes, relationships, and content at the nodes, feature engineering has presented a unique challenge for network mining tasks such as relation mining \cite{wang2010mining,sun2012will}, relevance search \cite{sun2011pathsim,huang2016meta}, personalized recommendation \cite{yu2014personalized,liu2014meta,ren2014cluscite}. The typical feature engineering activity responds to the requirements of the network mining task for an  application, requiring both a domain understanding and large exploratory search space for possible features. Not only this is expensive, but it also may not result in optimal performance. 
\begin{figure}
\begin{center}
\vspace{-0.06in}
\includegraphics[scale=0.46]{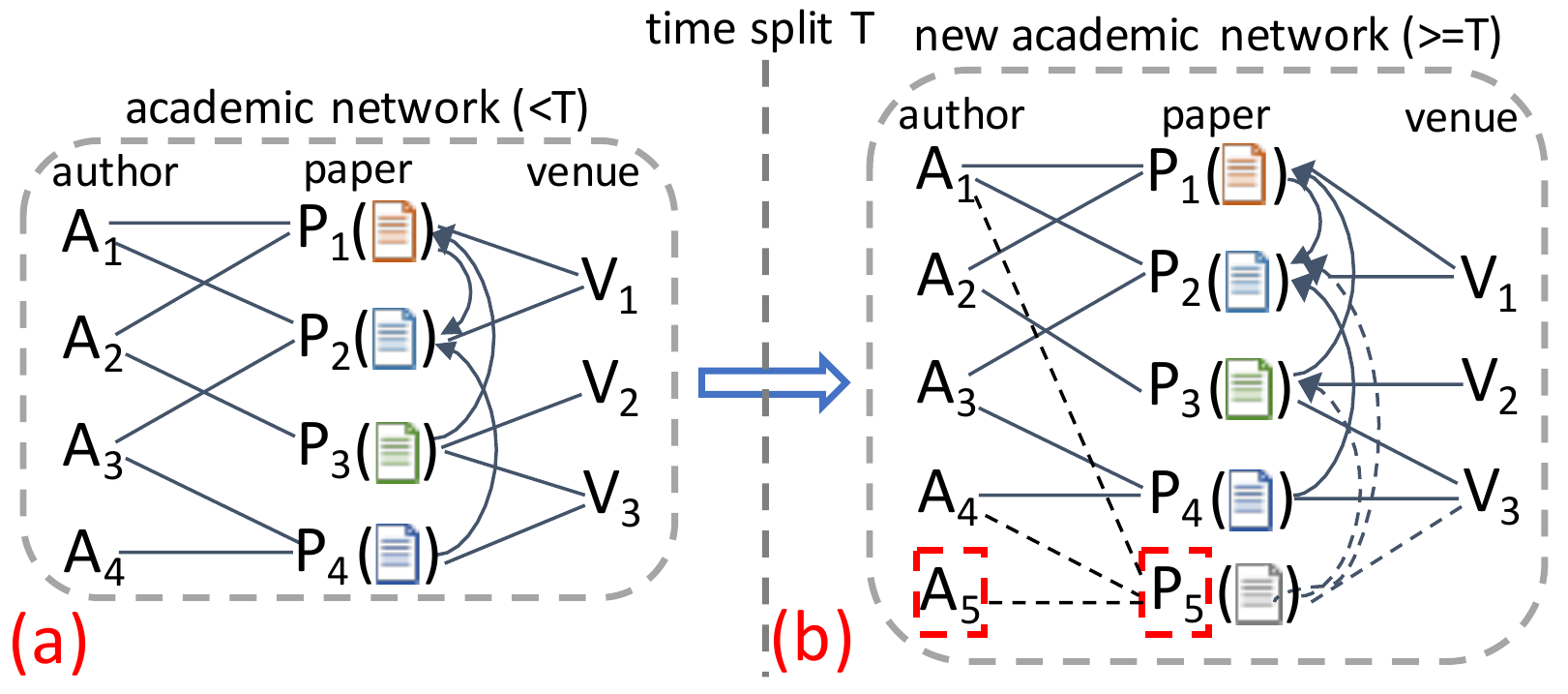}
 \vspace{-0.2in}
\caption{An Illustrative example of challenges in content-aware heterogeneous network representation learning.} 
 \vspace{-0.21in}
\label{fig: challenge}
\end{center}
\end{figure}

To that end, we ask the question: {\it{Can we generalize the feature engineering activity through representation learning on HetNets that addresses the complexity of multi-typed data in HetNets?}} With the advent of deep learning, significant effort has been devoted to network representation learning in the last few years, starting with a focus on homogeneous networks ~\cite{perozzi2014deepwalk, tang2015line, grover2016node2vec} and more recently on HetNets~\cite{dong2017metapath2vec}. 
The underlying theme of the models developed in these works is to automate the discovery of useful node latent features that can be further utilized in various network mining problems such as link prediction and node recommendation. However, these methods are limited in truly addressing the challenges of HetNets: 
\begin{itemize}[leftmargin=0.15in]\setlength{\itemsep}{0pt}
\item (\textbf{C1}) HetNets include multiple types of nodes and relations. For example, in Figure \ref{fig: challenge}(a), an academic network involves three types of nodes, i.e., \emph{author}, \emph{paper} and \emph{venue}, which are connected by three types of relations, i.e., \emph{author-write-paper}, \emph{paper-cite-paper} and \emph{paper-publish-venue}. Most of the previous models (e.g., Deepwalk and node2vec) employ homogeneous language models which make application to HetNets difficult. Thus challenge 1 is: how to extend homogeneous language model to heterogeneous network representation learning for maintaining structural closeness among multiple types of nodes and relations? We build on our prior work metapath2vec ~\cite{dong2017metapath2vec} for this.  

\item (\textbf{C2}) HetNets include both structural content (e.g., node type and relation connection) and unstructured semantic content (e.g., text). For example, in Figure \ref{fig: challenge}(a), paper in academic network connects to author \& venue and contains semantic text. The current models purely depend on structural content yet can not leverage unstructured content to infer semantic relations that are far away in network. To be more specific (as we will show in Section 4.5), given query author ``Jure Leskovec'', conventional techniques (e.g., node2vec and metapath2vec) tend to return authors who collaborated with ``Jure'' due to structural relations bridged by paper or return authors who are different from ``Jure'' in specific research interests due to structural relations bridged by venue. 
Thus challenge 2 is: how to effectively incorporate unstructured content of nodes into a representation learning framework for capturing both structural closeness and unstructured semantic relations among all nodes? 

\item (\textbf{C3}) HetNets can grow with time. Current models are not able to handle this due to lack of an update strategy and it is impractical to re-run the model for each new node. For example, in Figure \ref{fig: challenge}(b), new author $A_{5}$ co-authors with $A_{1}$ and $A_{4}$ on new paper $P_{5}$ after a given time split T. Thus challenge 3 is: how to efficiently learn representations of new nodes in a growing network? 
\end{itemize}

Our proposed method CARL, a {\bf \underline{c}}ontent-{\bf \underline{a}}ware {\bf \underline{r}}epresentation {\bf \underline{l}}earning model for HetNets, addresses these challenges. Specifically, 
first, we develop a heterogeneous SkipGram model to maintain structural closeness among multiple types of nodes and relations. Next, we design two effective ways based on deep semantic encoding to incorporate unstructured content (i.e., text) of some types of nodes into heterogeneous SkipGram for capturing semantic relations. The negative sampling technique and the walk sampling based strategy are utilized to optimize and train the proposed models. 
Finally, we develop an online update module to efficiently learn representation of each new node by using its relations with existing nodes and the learned node representations. 

To summarize, the main contributions of our work are:
\begin{itemize}[leftmargin=0.15in]\setlength{\itemsep}{0pt}
\item We formalize the problem of content-aware representation learning in HetNets and develop a model, i.e., CARL, to solve the problem. CARL performs joint optimization of heterogeneous SkipGram and deep semantic encoding. 

\item We design the corresponding optimization strategy and training algorithm to effectively learn node representations. The output representations are further utilized in various HetNet mining tasks, such as link prediction, document retrieval, node recommendation and relevance search, which demonstrate the superior performance of CARL over state-of-the-art baselines.

\item We propose an update module in CARL to handle growing networks and conduct the category visualization study to show the effectiveness of this module. 
\end{itemize}

\section{Problem Definition}\label{sec:problem}
\begin{figure*}
\begin{center}
\vspace{-0.1in}
\includegraphics[scale=0.48]{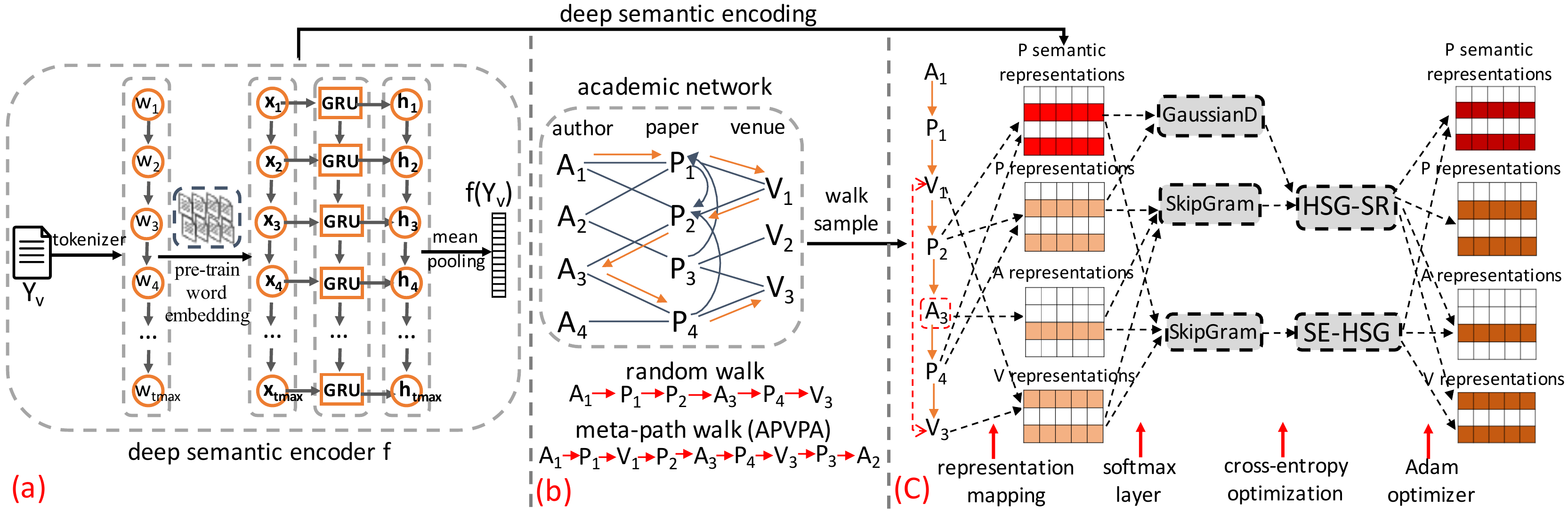}
\vspace{-0.18in}
\caption{An illustrative example of CARL in the academic network: (a) paper semantic encoder based on gated recurrent neural network; (b) academic network and random \& meta-path walks; (c) framework of the proposed models.} 
\vspace{-0.19in}
\label{fig: models}
\end{center}
\end{figure*}

We first introduce the concepts of HetNets and random \& meta-path walks, then formally define the problem of content-aware representation learning in HetNets. 
\begin{defn}
({\bf Heterogeneous Networks}) A heterogeneous network \cite{sun2012pathselclus} is defined as a network $G = (V, E, O_{V}, R_{E})$ with multiple types of nodes $V$ and links $E$. $O_{V}$ and $R_{E}$ represent the sets of object types and relation types. Each node $v \in V$ and each link $e \in E$ is associated with a node type mapping function $\psi_{v}: V \rightarrow O_{V}$ and a link type mapping function $\psi_{e}: E \rightarrow R_{E}$.
\end{defn}
For example, in Figure \ref{fig: models}(b), the academic network can be seen as a HetNet. The set of node types $O_{V}$ includes \emph{author} (A), \emph{paper} (P) and \emph{venue} (V). The set of link types $R_{E}$ includes \emph{author-write-paper}, \emph{paper-cite-paper} and \emph{paper-publish-venue}.

\begin{defn}
({\bf Random Walk}) A random walk \cite{grover2016node2vec} is defined as a node sequence $S_{v_{0}}=\{v_{0},v_{1},v_{2},...,v_{L-1}\}$ wherein the $i$-th node $v_{i-1}$ in the walk is randomly selected from the neighbors of its predecessor $v_{i-2}$.
\end{defn}

\begin{defn}
({\bf Meta-path Walk}) A meta-path walk \cite{dong2017metapath2vec} in HetNet is defined as a random walk guided by a specific meta-path scheme with the form of $\mathcal{P} \equiv o_{1}\overset{r_{1}}{\rightarrow}o_{2}\overset{r_{2}}{\rightarrow}\cdots \overset{r_{m-1}}{\rightarrow}o_{m}$, where $o_{i} \in O_{V}$, $r_{i} \in R_{E}$ and $r=r_{1} \ast r_{2} \cdots \ast r_{m-1}$ represents a compositional relation between relation types $r_{1}$ and $r_{m}$. Each meta-path walk recursively samples a specific $\mathcal{P}$ until it meets the given length. 
\end{defn}
Figure \ref{fig: models}(b) shows examples of random walk and ``APVPA'' meta-path walk in the academic network. 

\begin{defn}
({\bf Content-Aware Representation Learning in Heterogeneous Networks}) Given a HetNet with both structural and unstructured content at each node, the task is to design a model to learn a $d$-dimensional feature representations $\theta \in \mathbb{R}^{|V|\times d} (d \ll |V|)$, which can encode both structural closeness and unstructured semantic relations. Furthermore, the model is able to efficiently infer representation $\theta'_{v'} \in \mathbb{R}^{d}$ for each new node $v'$ by using the learned representations $\theta$. 
\end{defn}
For example, in the network of Figure \ref{fig: challenge}, author and venue nodes contain structural content, i.e., node id, node type as well as link relations with others, and paper node contains both structural content and unstructured semantic content, e.g., abstract text. 
The output $\theta$ denotes representations of all existing nodes via the same latent space, which can be further utilized in various HetNet mining tasks. Besides, the learned representation $\theta'_{v'}$ of each new node $v'$ can benefit different tasks for $v'$ such as category assignment. 


\section{CARL Framework}\label{sec:method}
We present the framework of content-aware representation learning which will address the three challenges described in Section 1. 

\subsection{Heterogeneous Network Embedding (\textbf{C1})}
Inspired by word2vec \cite{mikolov2013distributed} for learning distributed representation of words in corpus, Deepwalk \cite{perozzi2014deepwalk} and node2vec \cite{grover2016node2vec} leverage SkipGram and random walks to learn node representations. 
However, we argued that those techniques focus on homogeneous networks and proposed metapath2vec \cite{dong2017metapath2vec} for HetNets by feeding meta-path walks to SkipGram. Similar to metapath2vec, we formulate the heterogeneous network representation learning as heterogeneous SkipGram (HSG) to address challenge \textbf{C1}. Specifically, given a HetNet $G = (V, E, O_{V}, R_{E})$, the objective is to maximize the likelihood of each type of context node given the input node $v$: 
\begin{equation}
\begin{split}
\vspace{-0.2in}
&\mathlarger{o}_{1}=\arg\max_{\substack{\theta}}\prod_{v\in V}\prod_{t\in O_{V}}\prod_{v_{c}\in N_{t}(v)}p(v_{c}|v;\theta)
\vspace{-0.2in}
\end{split}
\label{equ: equ-2}
\end{equation}
where $\theta$ contains the representations of all nodes and $N_{t}(v)$ is the set of $t$-type context node of $v$ which can be collected in different ways such as one-hop neighbors or surrounding neighbors in random walks. For example, in Figure \ref{fig: models}(b), $A_{3}$ is structurally close to other authors (e.g., $A_{1}$ \& $A_{4}$), papers (e.g., $P_{2}$ \& $P_{4}$) and venues (e.g., $V_{1}$ \& $V_{3}$). Thus objective $\mathlarger{\mathlarger{o}}_{1}$ is able to maintain structural closeness among multiple types of nodes and relations in $G$. 

\subsection{Incorporating Deep Semantic Encoder (\textbf{C2})}
The objective $\mathlarger{\mathlarger{o}}_{1}$ formulates structural closeness but ignores unstructured semantic relations. To address challenge \textbf{C2}, we design two ways to incorporate unstructured content of some types of nodes into heterogeneous network representation learning. 

\subsubsection{HSG with Unstructured Semantic Regularization (HSG-SR)}
One way is to tightly join HSG with the conditional probability of semantic constraint, leading to unstructured semantic regularization onto objective $\mathlarger{\mathlarger{o}}_{1}$. The objective is defined as:
\begin{equation}
\begin{split}
\vspace{-0.2in}
&\mathlarger{o}_{2}=\arg\max_{\substack{\theta, \Phi}}\prod_{v\in V}\prod_{t\in O_{V}}\prod_{v_{c}\in N_{t}(v)}p(v_{c}|v;\theta)\prod_{v\in V_{S}}p(\theta_{v}|Y_{v};\Phi)
\vspace{-0.2in}
\end{split}
\label{equ: equ-5}
\end{equation}
where $V_{S}$ is the set of nodes with unstructured semantic content, $Y_{v}$ represents unstructured content of node $v$ and $\Phi$ are parameters of a deep semantic encoder that will be described later. The conditional probability $p(v_{c}|v;\theta)$ is defined as the heterogeneous softmax function: $p(v_{c}|v;\theta) = \frac{e^{\theta_{v_{c}}\cdot \theta_{v}}}{\sum_{v_{k}\in V_{t}}e^{\theta_{v_{k}}\cdot \theta_{v}}}$, where $V_{t}$ is the set of $t$-type nodes. Besides, we model the conditional probability $p(\theta_{v}|Y_{v};\Phi)$ as Gaussian prior: $p(\theta_{v}|Y_{v};\Phi) = {N}(\theta_{v}|E_{v}, \sigma^{2}{I})$, where $E_{v}$ denotes $v$'s semantic representation encoded by deep learning architecture $f$: $E_{v} = f(Y_{v})$. For example, in the network of Figure \ref{fig: models}, $V_{S}$ in $\mathlarger{\mathlarger{o}}_{2}$ is the set of papers and the formulation involves four kinds of representations, i.e., author representations, venue representations, paper representations and paper semantic representations. Notice that, there are two kinds of paper representations and we will use paper semantic representation for evaluation in Section 4. 


\subsubsection{Unstructured Semantic Enhanced HSG (SE-HSG)}
Another way is to concatenate the output of deep semantic encoder with the input of HSG, leading to unstructured semantic enhancement onto objective $\mathlarger{\mathlarger{o}}_{1}$. The objective is defined as: 
\begin{equation}
\begin{split}
\vspace{-0.2in}
&\mathlarger{o}_{3}=\arg\max_{\substack{\theta, \Phi}}\prod_{v\in V}\prod_{t\in O_{V}}\prod_{v_{c}\in N_{t}(v)}p(v_{c}|v;\theta;Y;\Phi)
\vspace{-0.2in}
\end{split}
\label{equ: equ-3}
\end{equation}
where $Y$ is the set of all unstructured semantic content of $V_{S}$. The conditional probability $p(v_{c}|v;\theta;Y;\Phi)$ is defined as the semantic enhanced heterogeneous softmax function: $p(v_{c}|v;\theta;Y;\Phi) = \frac{e^{\Theta_{v_{c}}\cdot \Theta_{v}}}{\sum_{v_{k}\in V_{t}}e^{\Theta_{v_{k}}\cdot \Theta_{v}}}$, where
$\Theta$ denotes the enhanced representations. That is, $\Theta_{v} = E_{v} = f(Y_{v})$ for $v \in V_{S}$ otherwise $\Theta_{v} =\theta_{v}$. For example, in the network of Figure \ref{fig: models}, $Y$ in $\mathlarger{\mathlarger{o}}_{3}$ is text content of all papers and the formulation involves three kinds of representations, i.e., author representations, venue representations and paper semantic representations. Notice that, $\theta$ in $\mathlarger{\mathlarger{o}}_{3}$ only denotes representations of nodes without unstructured content (e.g., author and venue in academic network), which is a bit different from $\mathlarger{\mathlarger{o}}_{2}$. 

\subsubsection{Unstructured Semantic Content Encoder.}
Both objectives $\mathlarger{\mathlarger{o}}_{2}$ and $\mathlarger{\mathlarger{o}}_{3}$ involve deep semantic encoding architecture. To encode unstructured content of some types of nodes into fixed length representations $E \in \mathbb{R}^{|V_{S}|\times d}$, we introduce gated recurrent units (GRU), a specific type of recurrent neural network, which has been widely adopted for many applications such as machine translation \cite{cho2014learning}. Figure \ref{fig: models}(a) gives an illustrative example of this encoder for papers in the academic network. To be more specific, each paper's abstract is represented as a sequence of words: $\{w_{1}, w_{2},  \cdots, w_{t_{max}}\}$, followed by the word embeddings sequence: $\{{\bf x}_{1}, {\bf x}_{2},  \cdots, {\bf x}_{t_{max}}\}$ trained by word2vec \cite{mikolov2013distributed}, where $t_{max}$ is the maximum length of text. For each step $t$ with the input word embedding ${\bf x}_{t}$ and previous hidden state vector ${\bf h}_{t-1}$, the current hidden state vector ${\bf h}_{t}$ is updated by ${\bf h}_{t} = {\bf GRU}({\bf x}_{t}, {\bf h}_{t-1})$, where the GRU module is defined as: 
\begin{equation}
\begin{split}
\vspace{-0.2in}
&{\bf z}_{t} = \sigma ({\bf A}_{z}{\bf x}_{t} + {\bf B}_{z}{\bf h}_{t-1}) \\
& {\bf r}_{t} = \sigma ({\bf A}_{r}{\bf x}_{t} + {\bf B}_{r}{\bf h}_{t-1}) \\
&{\bf \hat{h}}_{t} = \tanh [{\bf A}_{h}{\bf x}_{t} + {\bf B}_{h}( {\bf r}_{t}\circ {\bf h}_{t-1})]\\
& {\bf h}_{t} = {\bf z}_{t}\circ {\bf h}_{t-1} + (1-{\bf z}_{t})\circ {\bf \hat{h}}_{t} 
\vspace{-0.2in}
\end{split}
\label{equ: equ-11}
\end{equation}
where $\sigma$ is the sigmoid function, ${\bf A}$ and ${\bf B}$ are parameter matrices of GRU network (i.e., $\Phi$ in objectives $\mathlarger{\mathlarger{o}}_{2}$ and $\mathlarger{\mathlarger{o}}_{3}$ includes ${\bf A}$ and ${\bf B}$), operator $\circ$ denotes element-wise multiplication, ${\bf z}_{t}$ and ${\bf r}_{t}$ are update gate vector and reset gate vector, respectively. The GRU network encodes word embeddings to deep semantic embeddings ${\bf h} \in \mathbb{R}^{t_{max}\times d}$, which is concatenated with a mean pooling layer to obtain the general semantic representation of paper. All of these steps construct the deep semantic encoder $f$. We have also explored other encoding architectures such as LSTM, bidirectional GRU and attention-based GRU, and obtain similar results. Thus we choose GRU since it has a concise structure and reduce training time. 

\subsection{Model Optimization and Training}
We leverage the negative sampling technique \cite{mikolov2013distributed} to optimize model and introduce the walk sampling based strategy for model training.  

\subsubsection{Optimization of HSG-SR}
By applying negative sampling to the construction of softmax function, we can approximate the logarithm of $p(v_{c}|v;\theta)$ in objective $\mathlarger{\mathlarger{o}}_{2}$ as:
\begin{equation}
\begin{split}
\vspace{-0.2in}
& \log\sigma(\theta_{v_{c}}\cdot \theta_{v})+\sum_{m=1}^{M}\mathbb{E}_{v_{c'}\sim P_{t}(v_{c'})}\log\sigma(-\theta_{v_{c'}}\cdot \theta_{v})
\vspace{-0.2in}
\end{split}
\label{equ: equ-7}
\end{equation}
where $M$ is the negative sample size and $P_{t}(v_{c'})$ is the pre-defined sampling distribution w.r.t. the $t$-type node. In our case, $M$ makes little impact on the performance of proposed models. Thus we set $M=1$ and obtain the cross entropy loss for optimization:
\begin{equation}
\begin{split}
& \log p(v_{c}|v;\theta) = \log\sigma(\theta_{v_{c}}\cdot \theta_{v})+\log\sigma(-\theta_{v_{c'}}\cdot \theta_{v})
\end{split}
\label{equ: equ-9}
\end{equation}
That is, for each context node $v_{c}$ of $v$, we sample a negative node $v_{c'}$ according to $P_{t}(v_{c'})$. 
Besides, as $p(\theta_{v}|Y_{v};\Phi) = {N}(\theta_{v}|E_{v}, \sigma^{2}{I})$, the logarithm of $p(\theta_{v}|Y_{v};\Phi)$ in objective $\mathlarger{\mathlarger{o}}_{2}$ is equivalent to:
\begin{equation}
\begin{split}
& \log p(\theta_{v}|Y_{v};\Phi) =  -\big[\theta_{v}-E_{v}\big]^{T}\big[\theta_{v}-E_{v}\big]
\end{split}
\label{equ: equ-10}
\end{equation}
where $E_{v} = f(Y_{v})$ and $f$ is the deep semantic encoder. Therefore we rewrite objective $\mathlarger{\mathlarger{o}}_{2}$ as: 
\begin{equation}
\begin{split}
\mathlarger{o}_{2} =& \sum _{\left \langle v,v_{c},v_{c'} \right \rangle\in T_{walk}} \Big\{\log\sigma(\theta_{v_{c}}\cdot \theta_{v})+\log\sigma(-\theta_{v_{c'}}\cdot \theta_{v}) \\
&-\gamma \sum _{v_{*}\in T^{S}_{tri}} \big[\theta_{v_{*}}-f(Y_{v_{*}})\big]^{T}\big[\theta_{v_{*}}-f(Y_{v_{*}})\big]\Big\}
\end{split}
\label{equ: equ-12}
\end{equation}
where $\gamma$ is a trade-off factor and $T_{walk}$ denotes the set of triplets $\left \langle v,v_{c},v_{c'} \right \rangle$ collected by walk sampling on HetNet, which will be described later. Besides, $T^{S}_{tri}$ is the set of nodes with unstructured content in each triplet of $T_{walk}$ for semantic regularization. That is, $max\{|T^{S}_{tri}|\}=3$.

\subsubsection{Optimization of SE-HSG}
Similar to HSG-SR, the logarithm of $p(v_{c}|v;\theta;Y;\Phi)$ in objective $\mathlarger{\mathlarger{o}}_{3}$ is approximated by:
\begin{equation}
\begin{split}
\vspace{-0.2in}
&\log~p(v_{c}|v;\theta;Y;\Phi) = \log\sigma(\Theta_{v_{c}}\cdot \Theta_{v})+\log\sigma(-\Theta_{v_{c'}}\cdot \Theta_{v})
\end{split}
\label{equ: equ-8}
\end{equation}
where $\Theta_{v} = E_{v} = f(Y_{v})$ for nodes with unstructured content otherwise $\Theta_{v} =\theta_{v}$. Therefore we rewrite objective $\mathlarger{\mathlarger{o}}_{3}$ as: 
\begin{equation}
\begin{split}
&\mathlarger{o}_{3} = \sum _{\left \langle v,v_{c},v_{c'} \right \rangle\in T_{walk}} \log\sigma(\Theta_{v_{c}}\cdot \Theta_{v})+\log\sigma(-\Theta_{v_{c'}}\cdot \Theta_{v})
\end{split}
\label{equ: equ-13}
\end{equation}
As in HSG-SR, $T_{walk}$ is the set of triplets $\left \langle v,v_{c},v_{c'} \right \rangle$ collected from walk sequences on HetNet.  

\subsubsection{Model Training.} 
Both optimized objectives $\mathlarger{\mathlarger{o}}_{2}$ and $\mathlarger{\mathlarger{o}}_{3}$ are accumulated on set $T_{walk}$. Similar to Deepwalk, node2vec and metapath2vec, we design a walk sampling strategy to generate $T_{walk}$. Specifically, first, we uniformly generate a set of random walks or meta-path walks $S$ in HetNet. Then, for each node $v$ in $S_{i} \in S$, we collect context node $v_{c}$ which satisfies: $dist(v, v_{c})\leq \tau$. That is, $v$'s neighbors within distance $\tau$ in $S_{i}$. For example, in Figure \ref{fig: models}(c), the context node of $A_{3}$ ($\tau = 2$) in the sample walk are $V_{1}$, $P_{2}$, $P_{4}$ and $V_{3}$. Finally, for each $v_{c}$, we sample a negative node $v_{c'}$ with the same node type of $v_{c}$ according to $P_{t}(v_{c'}) \propto dg_{v_{c'}}^{3/4}$, where $dg_{v_{c'}}$ is the frequency of $v_{c'}$ in $S$. Furthermore, we design a mini-batch based Adam Optimizer \cite{kingma2014adam} to train the model. Specifically, at each iteration, we sample a mini-batch of triplets in $T_{walk}$ and accumulate the objective according to equation (8) or (10), then update the parameters via Adam. We repeat the training iterations until the change between two consecutive iterations is sufficiently small. Figure \ref{fig: models}(c) shows an illustration of the framework of HSG-SR and SE-HSG on the academic network. The output representations $\theta$ and $E$ can be utilized in various HetNet mining tasks, as we will show in Section 4. 

\subsection{Online Update for New Nodes (\textbf{C3})}
The optimization and training strategies in previous section learn representations of all existing nodes but they cannot be employed in an online situation. Considering the growing property of HetNets, we aim to design an online update module to efficiently learn representation of each new node and address challenge \textbf{C3}.
\begin{figure}
\begin{center}
\includegraphics[scale=0.39]{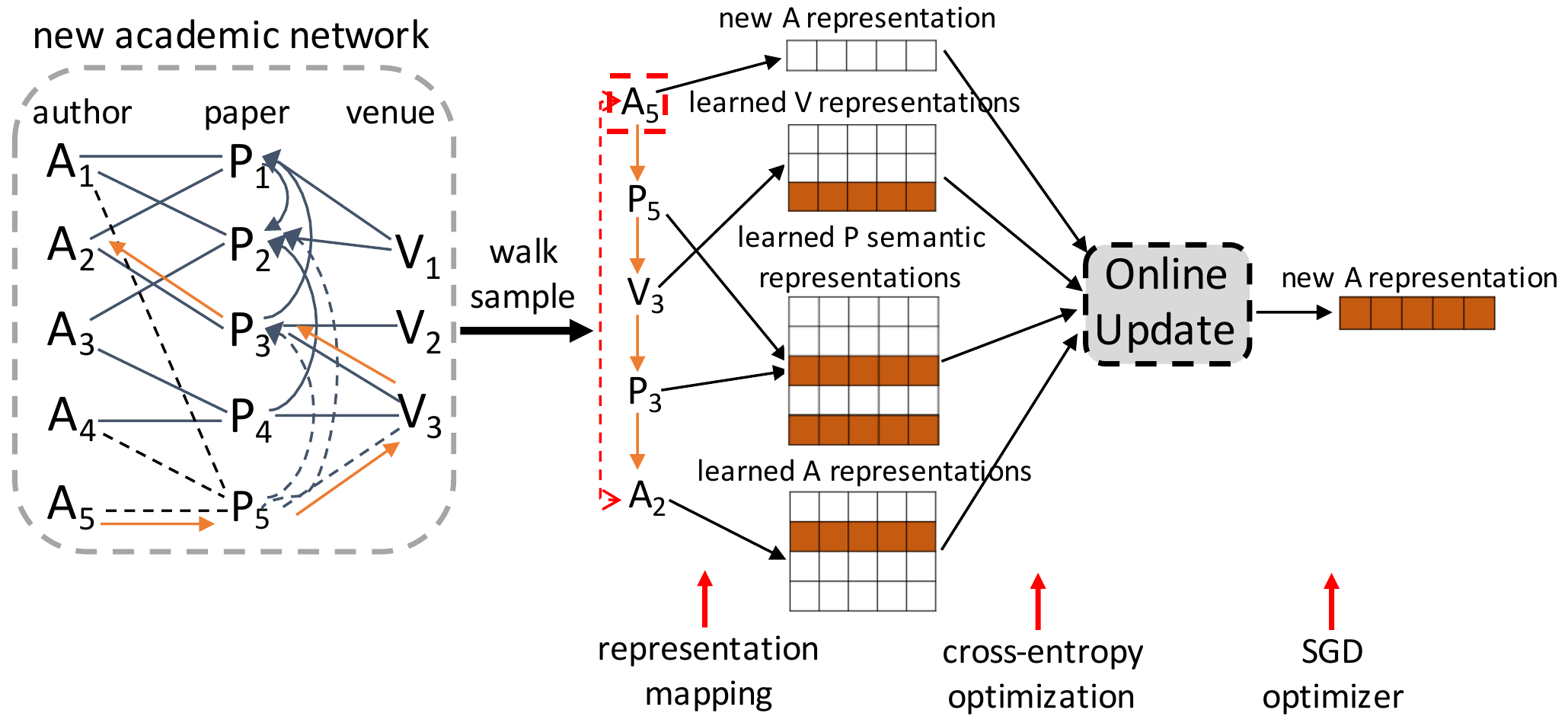}
\vspace{-0.17in}
\caption{An Illustrative example of online update for new author in academic network.} 
\vspace{-0.2in}
\label{fig: model-update}
\end{center}
\end{figure}

\subsubsection{Representation of New Node with Unstructured Semantic Content} 
As we introduce deep architecture $f$ to encode semantic representations of nodes with unstructured content, the optimized parameters $\Phi^{*}$ can be directly applied to infer representation of such new node without an extra training step. That is, $E_{v'} = f^{*}(Y_{v'})$ for new node $v'$, where $Y_{v'}$ is the text content of $v'$ and $f^{*}$ is the learned semantic encoder with $\Phi^{*}$. For example, in the academic network, we can use the learned paper semantic encoder to infer the representation of each new paper. 

\subsubsection{Representation of New Node without Unstructured Semantic Content} 
Inspired by eALS \cite{he2016fast} for online recommendation, we make a reasonable assumption that each new node should not change the learned node representations too much from a global perspective. Accordingly, for each incoming node $v'$ without unstructured content, we leverage objective $\mathlarger{\mathlarger{o}}_{3}$ to formulate the following objective:
\begin{equation}
\begin{split}
&\mathlarger{o}_{4} = \sum _{\left \langle v',v_{c},v_{c'} \right \rangle\in T^{v'}_{walk}} \log\sigma(\Theta^{*}_{v_{c}}\cdot \theta'_{v'})+\log\sigma(-\Theta^{*}_{v_{c'}}\cdot \theta'_{v'})
\end{split}
\label{equ: equ-14}
\end{equation}
where $\theta'$ denotes representation of $v'$, $\Theta^{*}$ is the learned node representations via SE-HSG, i.e., $\Theta^{*}_{v}=E^{*}_{v} = f^{*}(Y_{v})$ for $v \in V_{S}$ otherwise $\Theta^{*}_{v}=\theta^{*}_{v}$.
That is, we take $\Theta^{*}$ as constant values in $\mathlarger{\mathlarger{o}}_{4}$. In addition, we leverage meta-path walks rooted at $v'$ to collect triplet set $T^{v'}_{walk}$ for the training process of update module. Specifically, Figure \ref{fig: model-update} gives an illustrative example of online update for new author in academic network. First, we randomly sample a number of ``APVPA'' meta-path walks rooted at new author node $A_{5}$. The length of each walk equals the window distance $\tau$ ($\tau = 4$) and nodes in the walk are context node of $A_{5}$, e.g., $P_{5}$, $V_{3}$, $P_{3}$ and $A_{2}$. For each context node $v_{c}$, we randomly sample a negative node $v_{c'}$ with the same node type as $v_{c}$. Furthermore, the SGD optimizer is utilized to repeatly update $\theta'$ for each triplet in $T^{v'}_{walk}$ until the change between two consecutive iterations is sufficiently small. The proposed update module is efficient since only simple meta-path walk sampling and limited update steps on new node representation are performed. In the experimental evaluation, we only consider new author who writes new paper at a venue that exists in network, making the ``APVPA'' meta-path walk always feasible.  



\section{Experiments}\label{sec:experiments}
In this section, we conduct extensive experiments with the aim of answering the following research questions: 
\begin{itemize}[leftmargin=0.15in]\setlength{\itemsep}{0pt}
\item (\textbf{RQ1}) How does CARL perform vs. state-of-the-art network representation learning models for different HetNet mining tasks, such as link prediction (\textbf{RQ1-1}), document retrieval (\textbf{RQ1-2}) and node recommendation (\textbf{RQ1-3})? In addition, how do hyper-parameters impact CARL's performance in each task?
\item (\textbf{RQ2}) What is the performance difference between CARL and baselines in relevance search w.r.t. each task in \textbf{RQ1}? 
\item (\textbf{RQ3}) What is the performance of CARL's online update module on the task for new nodes such as category assignment?
\end{itemize}

Notice that, although our model can be applied to or modified for different HetNets, we focus on experiments on the academic HetNet due to data availability. 
\subsection{Experimental Setup}

\subsubsection{Data}
We use the AMiner computer science dataset \cite{tang2008arnetminer}, which is publicly available\footnote{https://aminer.org/citation}. To avoid noise, we remove the papers published in venues (e.g., workshop) with limited publications and the instances without abstract text. In addition, topic of each area changes over time. For example, according to our analysis of the data, 
the most popular topics in data mining change from web mining and clustering (1996$\sim$2005) to network mining and learning (2006$\sim$2015). To make a thorough evaluation of CARL and verify its effectiveness for networks in different decades, we independently conduct experiments on two datasets, i.e., AMiner-$\uppercase\expandafter{\romannumeral1}$ (1996$\sim$2005) and AMiner-$\uppercase\expandafter{\romannumeral2}$ (2006$\sim$2015). As a result,  AMiner-$\uppercase\expandafter{\romannumeral1}$ contains 160,713 authors, 111,409 papers and 150 venues, AMiner-$\uppercase\expandafter{\romannumeral2}$ contains 571,693 authors, 483,449 papers and 492 venues. The structure of the academic network used in this work is shown in Figure \ref{fig: challenge}. 



\subsubsection{Comparison Baselines} 
We compare CARL with four state-of-the-art models, i.e., Deepwalk \cite{perozzi2014deepwalk}, LINE \cite{tang2015line}, node2vec \cite{grover2016node2vec} and metapath2vec \cite{dong2017metapath2vec}. 
Notice that, we use either random walk (rw) or meta-path walk (mw) to collect context node in HSG-SR and SE-HSG, resulting in four variants of CARL: CARL$^{rw}_{HSG-SR}$, CARL$^{mw}_{HSG-SR}$, CARL$^{rw}_{SE-HSG}$ and CARL$^{mw}_{SE-HSG}$. 

\subsubsection{Reproducibility}
For fairness comparison, we use the same representation dimension $d$ = 128 for all models. The window size $\tau$ = 7, the number of walks per node $N$ = 10 and the walk length $L$ = 30 are used for Deepwalk, node2vec, metapath2vec and CARL. The same set of parameters is used for CARL's online update module. The size of negative samples $M$ is set to 5 for node2vec, LINE and metapath2vec. In addition, $\gamma =1.0$ for CARL$_{HSG-SR}$ and three meta-path schemes "APA", "APPA" and "APVPA" are jointly used to generate meta-path walks for CARL$^{mw}$. We employ TensorFlow to implement all variants of CARL and further conduct them on NVIDIA TITAN X GPU. Code will be available upon publication. 


\subsection{Link Prediction (\textbf{RQ1-1})}
\textbf{Who will be your academic collaborators?} As a response to \textbf{RQ1-1}, we design an experiment to evaluate CARL's performance on the author collaboration link prediction task. 
\subsubsection{Experimental Setting}
Unlike past work \cite{grover2016node2vec} that randomly samples a portion of links for training and uses the remaining for evaluation, we consider a more realistic setting that splits training/test data via a given time stamp T. Specifically, first, the network before T is utilized to learn node representations. Then, the collaboration links before T are used to train a binary logistic classifier. Finally, the collaboration relations after T with equal number of random non-collaboration links are used to evaluate the trained classifier. In addition, only new collaborations among current authors (who appear before T) are considered and duplicated collaborations are removed from evaluation. For example, in Figure \ref{fig: challenge}(b), $A_{1}$ and $A_{4}$ co-author a new paper $P_{5}$ after T. The classifier tends to predict new collaboration between $A_{1}$ and $A_{4}$ using previous links, e.g., collaboration between $A_{1}$ and $A_{3}$. The representation of link is formed by element-wise multiplication between representations of two end nodes. We use \textbf{Accuracy} and \textbf{F1 score} of binary classification as the evaluation metrics. Besides, T is set as 2003/2004 and 2013/2014 for AMiner-$\uppercase\expandafter{\romannumeral1}$ and AMiner-$\uppercase\expandafter{\romannumeral2}$, respectively. 

\subsubsection{Results}
The performances of different models are reported in Table \ref{tab: result-AA}. According to the table: (a) All variants of CARL perform better than baselines, demonstrating the effectiveness of incorporating unstructured semantic content to learn author representations; (b) CARL$^{mw}_{SE-HSG}$ achieves the best performances in all cases. The average improvements of CARL$^{mw}_{SE-HSG}$ over different baselines range from 10.9\% to 41.0\% and 6.7\% to 30.9\% on AMiner-$\uppercase\expandafter{\romannumeral1}$ and AMiner-$\uppercase\expandafter{\romannumeral2}$, respectively. (c) CARL$^{mw}$ outperforms CARL$^{rw}$, showing that meta-path walk is better than random walk for collecting context node in CARL. In addition, CARL$_{SE-HSG}$ has better performance than CARL$_{HSG-SR}$, indicating that concatenating text encoder with heterogeneous SkipGram is more significant than taking text encoding as semantic regularization. 

\begin{table}[!tb]
\caption{Collaboration prediction results comparison.} 
\vspace{-0.18in}
\begin{center}
\small
\begin{tabular}{c||c|c|c|c|c}
  \toprule
   \multirow{2}{*}{AMiner-$\uppercase\expandafter{\romannumeral1}$}&  \multicolumn{2}{c|}{T = 2004} &  \multicolumn{2}{c|}{T = 2003} & \multirow{2}{*}{Gain}\\
      \cmidrule{2-5}  
  & Accuracy &  F1 & Accuracy &  F1&\\
    \midrule
    \midrule
 Deepwalk & 0.6341& 0.4323&0.6244 &0.4058 & 41.0\%\\
    LINE &0.6722 &0.5263 &0.6714 & 0.5231 & 20.8\%\\
  node2vec & 0.6758&0.5291 & 0.6821&0.5409 & 18.9\%\\
  metapath2vec&0.7013* &0.5914*&0.7041* &0.5935* &10.9 \%\\
   \midrule
  CARL$^{rw}_{HSG-SR}$& 0.7302 &0.6561 & 0.7378 & 0.6623&--\\
    CARL$^{mw}_{HSG-SR}$& 0.7367 & 0.6618 & 0.7401 &0.6635  &--\\
 CARL$^{rw}_{SE-HSG}$& 0.7388& 0.6579&0.7419 &0.6648&--\\
 CARL$^{mw}_{SE-HSG}$& \bf{0.7482} & \bf{0.6753} &\bf{0.7525} & \bf{0.6881} &--\\
   \midrule
   \midrule
\multirow{2}{*}{AMiner-$\uppercase\expandafter{\romannumeral2}$}&  \multicolumn{2}{c|}{T = 2014} &  \multicolumn{2}{c|}{T = 2013} & \multirow{2}{*}{Gain}\\
      \cmidrule{2-5}  
  & Accuracy &  F1 & Accuracy &  F1&\\
    \midrule
    \midrule
 Deepwalk & 0.6559& 0.5024&0.6487 &0.4833 &30.9\%\\
    LINE &0.7034 &0.6048 &0.6956 & 0.5898 &14.3\%\\
  node2vec & 0.7136&0.6122 & 0.7066&0.5965 &12.7\%\\
  metapath2vec&0.7299* & 0.6628*&0.7254* & 0.6512* &6.7\%\\
   \midrule
  CARL$^{rw}_{HSG-SR}$&0.7498 &0.7061  &0.7495 &0.6946 &--\\
    CARL$^{mw}_{HSG-SR}$&0.7511 &0.7084 &0.7503 &0.6962 &--\\
 CARL$^{rw}_{SE-HSG}$&0.7562 & 0.7125&0.7546 &0.6978&--\\
 CARL$^{mw}_{SE-HSG}$&\bf{0.7627} &\bf{0.7208} &\bf{0.7602} &\bf{0.7097} &--\\
   \midrule
  \end{tabular}
\end{center}
\vspace{-0.31in}
\label{tab: result-AA}
\end{table}%

\subsubsection{Parameter Sensitivity}
\begin{figure}
\begin{center}
     \subfigure[][Window size $\tau$]{
        \centering
        \includegraphics[width=3.9cm,height=2.4cm]{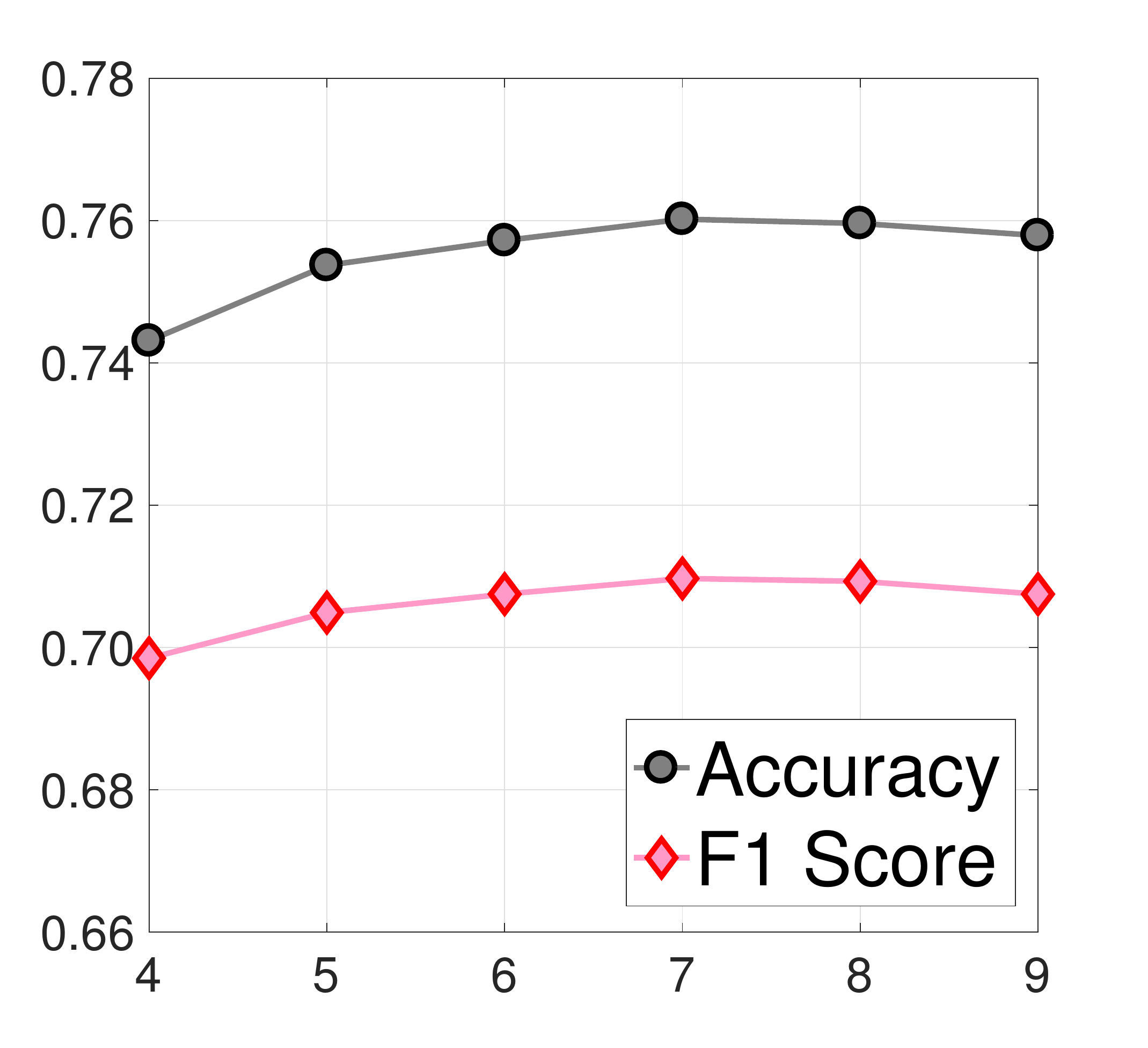}
        \vspace{-0.2in}
        \label{fig: impacts-AA-window}
        }
    \subfigure[][Representation dimension $d$]{
        \centering
        \includegraphics[width=3.9cm,height=2.4cm]{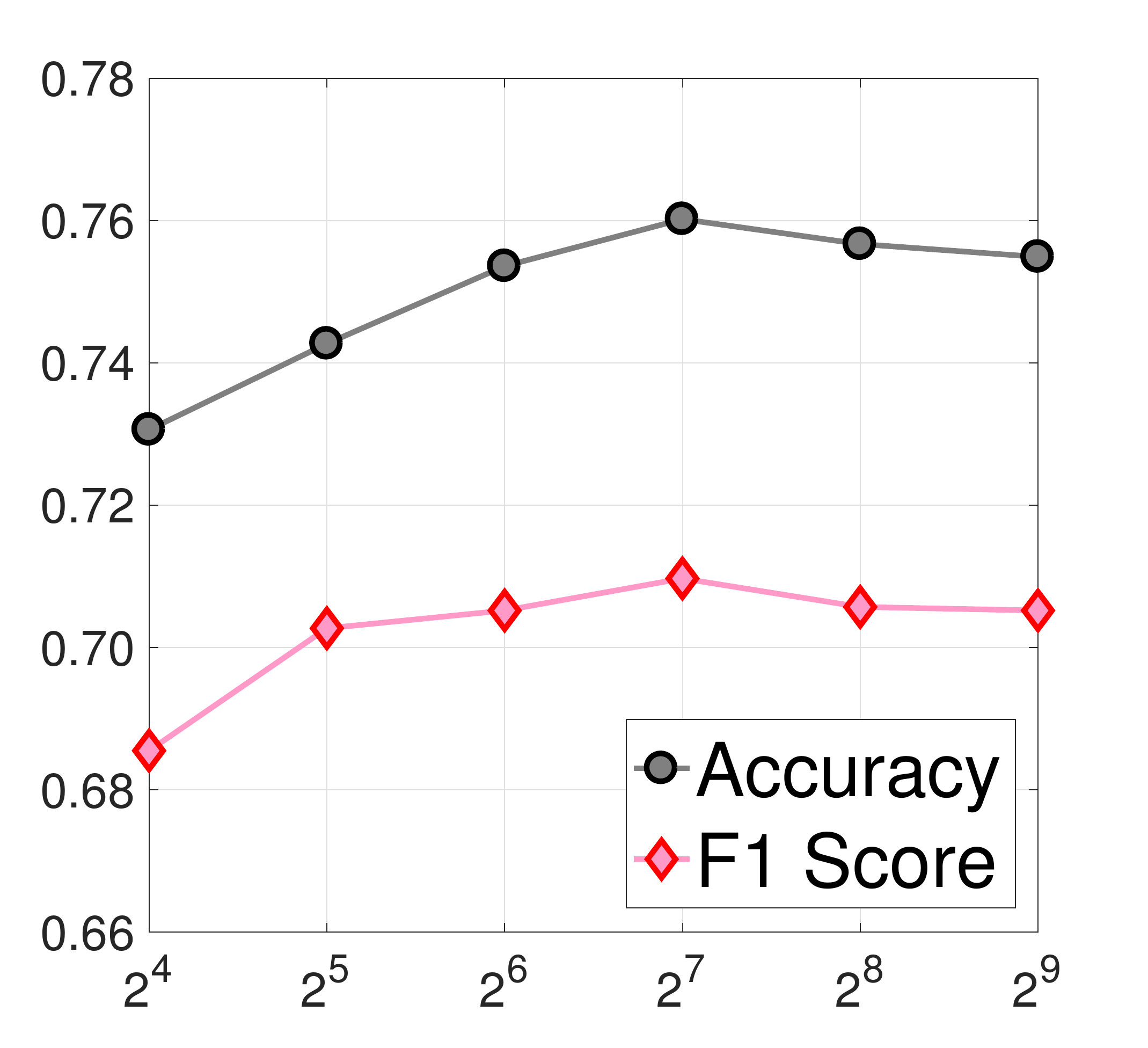}
        \vspace{-0.2in}
        \label{fig: impacts-AA-d}
        }
        \vspace{-0.2in}
    \caption{Parameter sensitivity in collaboration prediction.} 
     \vspace{-0.2in}
    \label{fig: impacts-AA}
\end{center}
\end{figure}

We conduct experiment to analyze the impact of two key parameters, i.e., the window size $\tau$ of walk sampling and the representation dimension $d$. We investigate a specific parameter by changing its value and fixing the others. The prediction results of CARL$^{mw}_{SE-HSG}$ as a function of $\tau$ and $d$ on AMiner-$\uppercase\expandafter{\romannumeral2}$ (T = 2013) are shown in Figure \ref{fig: impacts-AA}. We see that: (a) With increasing of $\tau$, accuracy and F1 score increase at first since a larger window means more useful context information. But when $\tau$ goes beyond a certain value, performances decrease slowly with $\tau$ possibly due to uncorrelated noise. The best $\tau$ is around 7; (b) Similar to $\tau$, an appropriate value should be set for $d$ such that the best node representations are learned. The optimal $d$ is around 128.

\subsection{Document Retrieval (\textbf{RQ1-2})}
\textbf{Which relevant papers should be retrieved for your query?} As a response to \textbf{RQ1-2}, we design an experiment to evaluate CARL's performance on the paper retrieval task. 

\subsubsection{Experimental Setting}
As in the previous task, the network before T is utilized to learn node representations. The ground truth of relevance is assumed as the co-cited relation between two papers after T. For example, in Figure \ref{fig: challenge}(b), new paper $P_{5}$ cites both previous papers $P_{2}$ and $P_{3}$. The model tends to retrieve $P_{2}$ when querying $P_{3}$ or retrieve $P_{3}$ when querying $P_{2}$. 
The relevant score of two papers is defined as the cosine similarity between representations of two papers. We use \textbf{HitRatio@k} as the evaluation metric. Due to large number of candidate papers, we follow the sampling strategy in \cite{yang2017bridging} to reduce evaluation time. Specifically, for each evaluated paper, we randomly generate 100 negative samples for comparison with the true relevant paper. The hit ratio equals 1 if the true relevant paper is ranked in the top-k list of relevant score, otherwise 0. The overall result is the average value of HitRatio@k among all evaluated papers. The duplicated co-cited relations are removed from evaluation and k is set to 10 or 20. In addition, T is set as 2003/2004 and 2013/2014 for AMiner-$\uppercase\expandafter{\romannumeral1}$ and AMiner-$\uppercase\expandafter{\romannumeral2}$, respectively. 
\subsubsection{Results}
The results are reported in Table \ref{tab: result-PP}. From the table: (a) All variants of CARL achieve better performance than baselines, demonstrating the benefit of incorporating unstructured semantic content to learn paper representations; (b) The average improvements of CARL$^{mw}_{SE-HSG}$ over different baselines range from 2.3\% to 16.4\% and 6.3\% to 29.1\% on AMiner-$\uppercase\expandafter{\romannumeral1}$ and AMiner-$\uppercase\expandafter{\romannumeral2}$, respectively; (c) CARL$_{SE-HSG}$ outperforms CARL$_{HSG-SR}$, showing that semantic enhanced SkipGram is better than SkipGram with semantic regularization. However, CARL$^{rw}$ has performance close to that of CARL$^{mw}$, indicating that meta-path walk has little impact for this task. It is reasonable since the paper representations depend on the deep semantic encoder in our model. 

 \begin{table}[!tb]
\caption{Paper retrieval results comparison.} 
\vspace{-0.18in}
\begin{center}
\small
\begin{tabular}{c||c|c|c|c|c}
  \toprule
   \multirow{2}{*}{AMiner-$\uppercase\expandafter{\romannumeral1}$}&  \multicolumn{2}{c|}{T = 2004} &  \multicolumn{2}{c|}{T = 2003} &\multirow{2}{*}{Gain}\\
      \cmidrule{2-5}  
  & Hit@10 &  Hit@20& Hit@10 &  Hit@20&\\
    \midrule
        \midrule
 Deepwalk &0.8120 &0.8816 &0.8217 &0.8967& 6.7\%\\
    LINE & 0.7320&0.8130 & 0.7485&0.8380& 16.4\%\\
  node2vec &0.8552* &0.9148* & 0.8653* & 0.9250* &2.3\%\\
  metapath2vec&0.8239&0.8910& 0.8366&0.9081 & 5.3\%\\
   \midrule
   CARL$^{rw}_{HSG-SR}$&0.8685  & 0.9351 &0.8696 &0.9375 &-- \\
  CARL$^{mw}_{HSG-SR}$&0.8673 & 0.9342& 0.8722&0.9389 &--\\
  CARL$^{rw}_{SE-HSG}$&0.8741 &0.9412  &\bf{0.8788} &0.9455 &--\\
  CARL$^{mw}_{SE-HSG}$&\bf{0.8751} &\bf{0.9425} & 0.8783 & \bf{0.9470} & --\\
   \midrule
    \midrule
\multirow{2}{*}{AMiner-$\uppercase\expandafter{\romannumeral2}$}&  \multicolumn{2}{c|}{T = 2014} &  \multicolumn{2}{c|}{T = 2013} &\multirow{2}{*}{Gain}\\
      \cmidrule{2-5}  
  & Hit@10 &  Hit@20& Hit@10 &  Hit@20&\\
    \midrule
        \midrule
 Deepwalk & 0.7460 &0.8366 &0.7392 &0.8316& 13.2\%\\
    LINE & 0.6502&0.7453 & 0.6353&0.7382& 29.1\%\\
  node2vec &0.8041*& 0.8785*&0.7981*&0.8749*& 6.3\%\\
  metapath2vec&0.7214 &0.8136 & 0.7257&0.8201 & 15.9\%\\
   \midrule
   CARL$^{rw}_{HSG-SR}$& 0.8263 & 0.9162 &0.8141  & 0.9101&-- \\
  CARL$^{mw}_{HSG-SR}$& 0.8287  & 0.9184 & 0.8153& 0.9096 &--\\
  CARL$^{rw}_{SE-HSG}$& 0.8619& \bf{0.9281} &0.8516 &0.9243 &--\\
  CARL$^{mw}_{SE-HSG}$& \bf{0.8625}&0.9278 &\bf{0.8518}  &\bf{0.9254}& --\\
   \midrule
   \end{tabular}
\end{center}
\vspace{-0.31in}
\label{tab: result-PP}
\end{table}%

\subsubsection{Parameter Sensitivity}
\begin{figure}
\begin{center}
     \subfigure[][Window size $\tau$]{
        \centering
        \includegraphics[width=3.9cm,height=2.4cm]{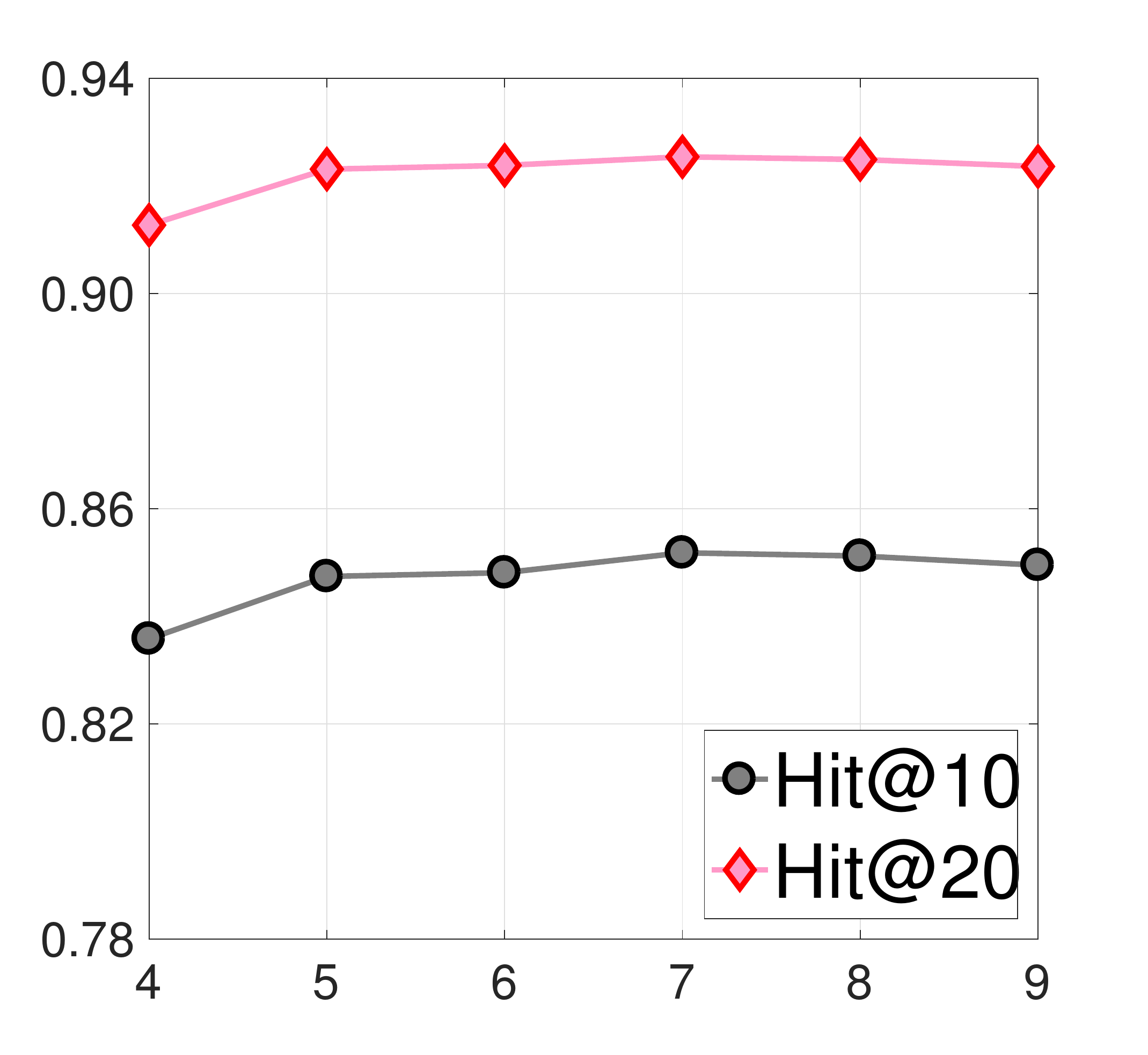}
        \vspace{-0.2in}
        \label{fig: impacts-PP-window}
        }
    \subfigure[][Representation dimension $d$]{
        \centering
        \includegraphics[width=3.9cm,height=2.4cm]{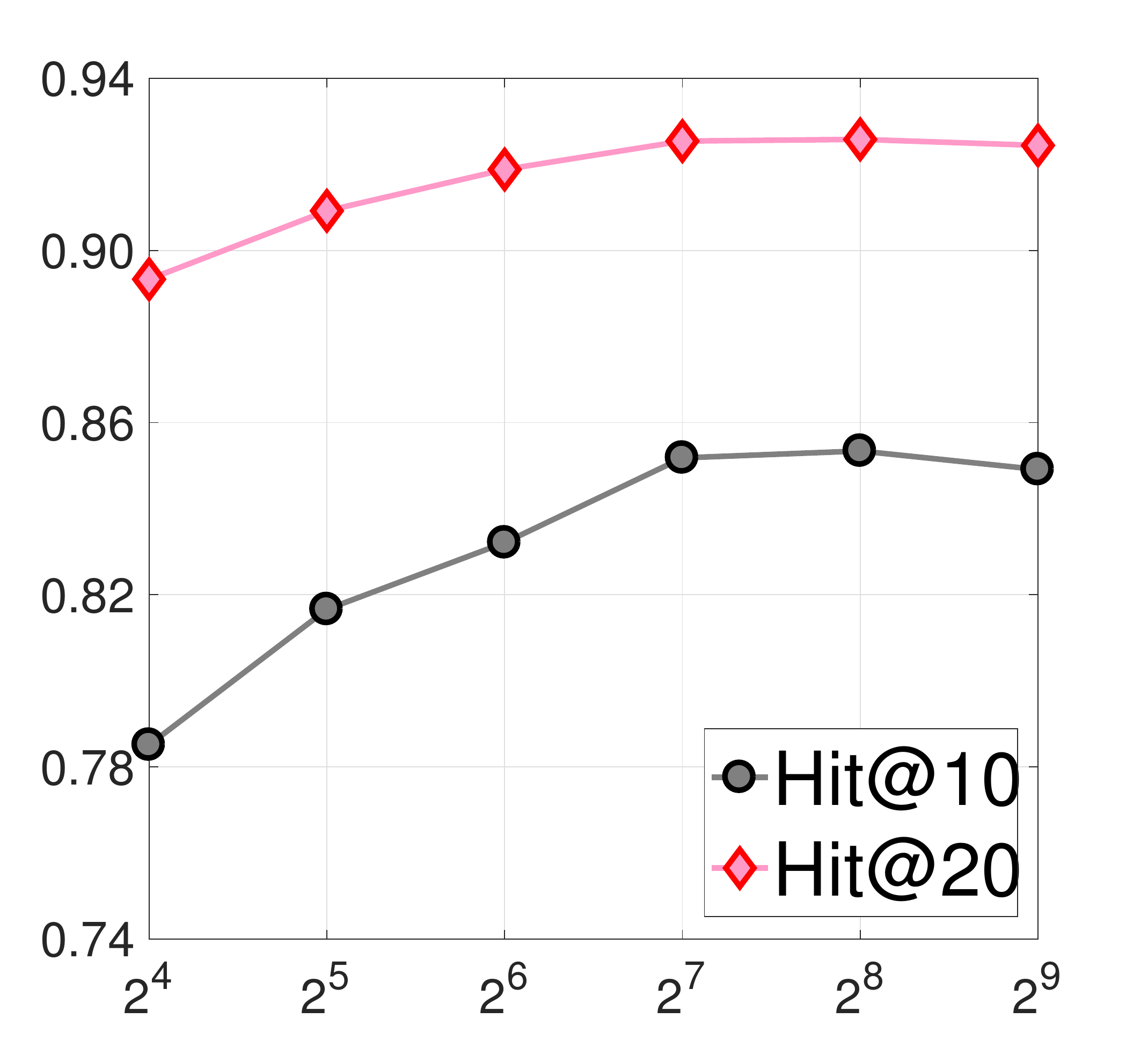}
        \vspace{-0.2in}
        \label{fig: impacts-PP-d}
        }
        \vspace{-0.2in}
    \caption{Parameter sensitivity in paper retrieval.}
     \vspace{-0.2in}
    \label{fig: impacts-PP}
\end{center}
\end{figure}

Following the same setup in link prediction, we investigate the impact of window size $\tau$ and representation dimension $d$ on CARL$^{mw}_{SE-HSG}$'s performance on AMiner-$\uppercase\expandafter{\romannumeral2}$ (T = 2013), as shown by Figure \ref{fig: impacts-PP}. It can be seen that: (a) The results are little sensitive to $\tau$ when $\tau \geq7$. As we noted above, paper representations depend on the deep semantic encoder; (b) The dimension $d$ plays significant role on generating paper representations. The best representations are learned when $d$ is around 128 for the paper retrieval task.   

\subsection{Node Recommendation (\textbf{RQ1-3})}
\textbf{Which venues should be recommended to you?} As a response to \textbf{RQ1-3}, we design an experiment to evaluate CARL's performance on the venue recommendation task. 

\subsubsection{Experimental Setting}
As in the previous two tasks, the network before T is utilized to learn node representations. The ground truth of recommendation is based on author's appearance in venue after T. 
For example, in Figure \ref{fig: challenge}(b), author $A_{1}$ writes new paper $P_{5}$ on $V_{3}$ after T, indicating $A_{1}$ is likely to accept $V_{3}$ recommendation before T. The preference score is defined as the cosine similarity between representations of author and venue. We use \textbf{Recall@k} as the evaluation metric and k is set to 5 or 10. In addition, the duplicated author-venue pairs are removed from evaluation. The reported score is the average value over all evaluated authors.  

\begin{table}[!tb]
\caption{Venue recommendation results comparison.} 
\vspace{-0.18in}
\begin{center}
\small
\begin{tabular}{c||c|c|c|c|c}
  \toprule
   \multirow{2}{*}{AMiner-$\uppercase\expandafter{\romannumeral1}$}&  \multicolumn{2}{c|}{T = 2004} &  \multicolumn{2}{c|}{T = 2003} & \multirow{2}{*}{Gain}\\
      \cmidrule{2-5}  
 & Rec@5 &  Rec@10 & Rec@5 &  Rec@10 & \\
    \midrule
    \midrule
 Deepwalk & 0.1051* &0.1628* &0.0864*& 0.1403*& 18.9\%\\
    LINE &0.0376 &0.0677 &0.0388 &0.0717 & 178.7\%\\
  node2vec &0.0945  &0.1570 &0.0774 &0.1386 & 27.2\%\\
  metapath2vec&0.0878 &0.1527 &0.0714 &0.1395 & 33.3\%\\
   \midrule
 CARL$^{rw}_{HSG-SR}$&0.1156  & 0.1772 &0.0988  &0.1593  &--\\
   CARL$^{mw}_{HSG-SR}$& 0.1208 & 0.1813&0.1050 &0.1631  &--\\
  CARL$^{rw}_{SE-HSG}$&0.1225 &0.1831 &0.1054 &0.1648&--\\
     CARL$^{mw}_{SE-HSG}$&\bf{0.1250} &\bf{0.1852} & \bf{0.1073}& \bf{0.1667}&--\\
   \midrule
   \midrule
\multirow{2}{*}{AMiner-$\uppercase\expandafter{\romannumeral2}$}&  \multicolumn{2}{c|}{T = 2014} &  \multicolumn{2}{c|}{T = 2013} & \multirow{2}{*}{Gain}\\
      \cmidrule{2-5}  
 & Rec@5 &  Rec@10 & Rec@5 &  Rec@10 & \\
    \midrule
      \midrule
 Deepwalk & 0.1039* &0.1549* &0.0925*& 0.1384*&11.7\%\\
    LINE &0.0297  &0.0470&0.0267 &0.0433&275.6\%\\
  node2vec &0.0766  &0.1182 &0.0691 &0.1074&47.8\%\\
  metapath2vec& 0.0654 &0.1077&0.0608 &0.1001&66.1\%\\
   \midrule
 CARL$^{rw}_{HSG-SR}$&0.1092  &0.1643 &0.0977 &0.1466&--\\
   CARL$^{mw}_{HSG-SR}$&0.1116  &0.1713 &0.1032 &0.1526&--\\
  CARL$^{rw}_{SE-HSG}$& 0.1107 &0.1725 &0.1023 &0.1534&--\\
     CARL$^{mw}_{SE-HSG}$& \bf{0.1136} & \bf{0.1742} &\bf{0.1045} &\bf{0.1551}&--\\
   \midrule
   \end{tabular}
\end{center}
\vspace{-0.31in}
\label{tab: result-AV}
\end{table}%

\begin{figure}
\begin{center}
     \subfigure[][Window size $\tau$]{
        \centering
        \includegraphics[width=3.9cm,height=2.4cm]{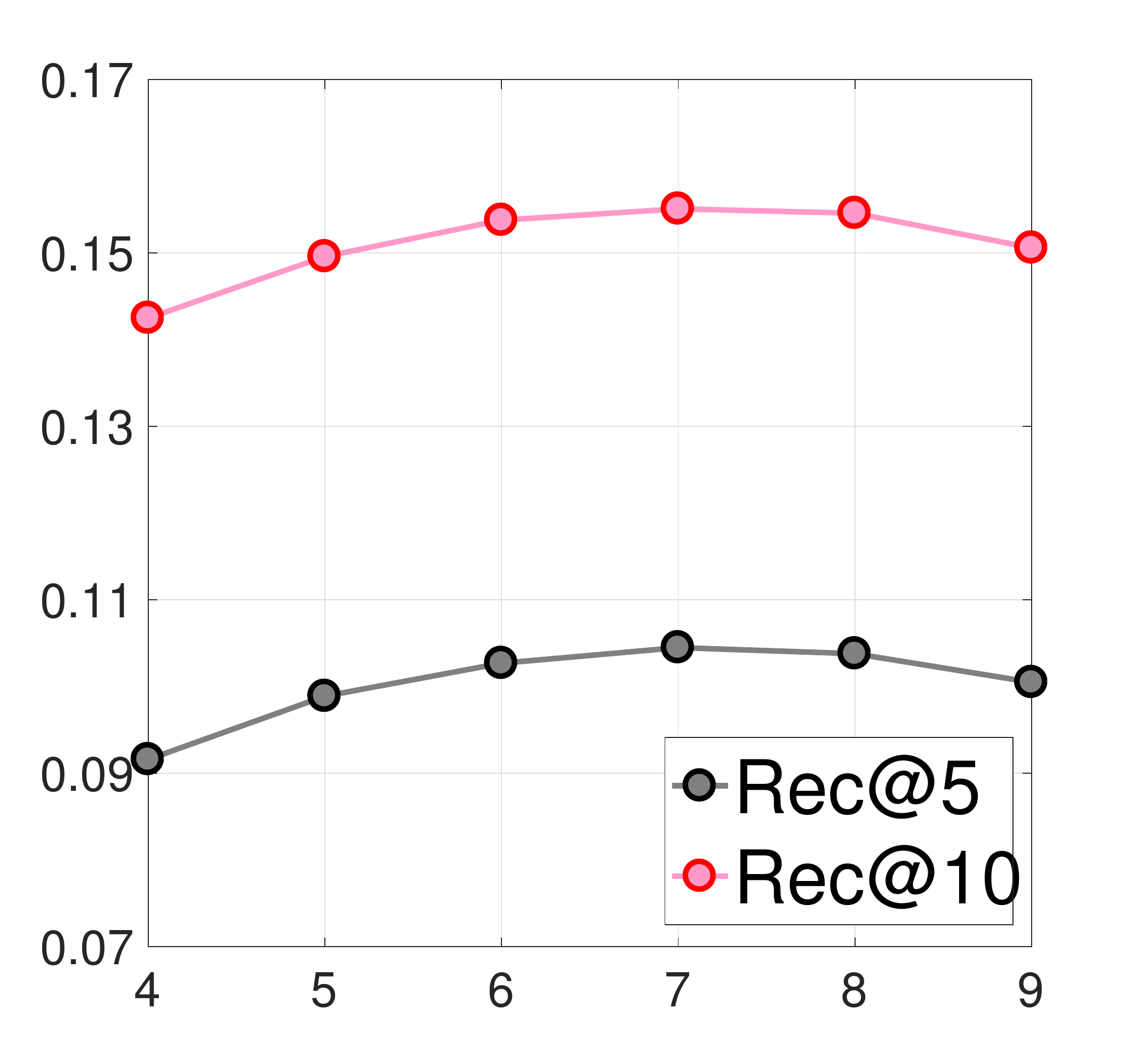}
        \vspace{-0.2in}
        \label{fig: impacts-AV-window}
        }
    \subfigure[][Representation dimension $d$]{
        \centering
        \includegraphics[width=3.9cm,height=2.4cm]{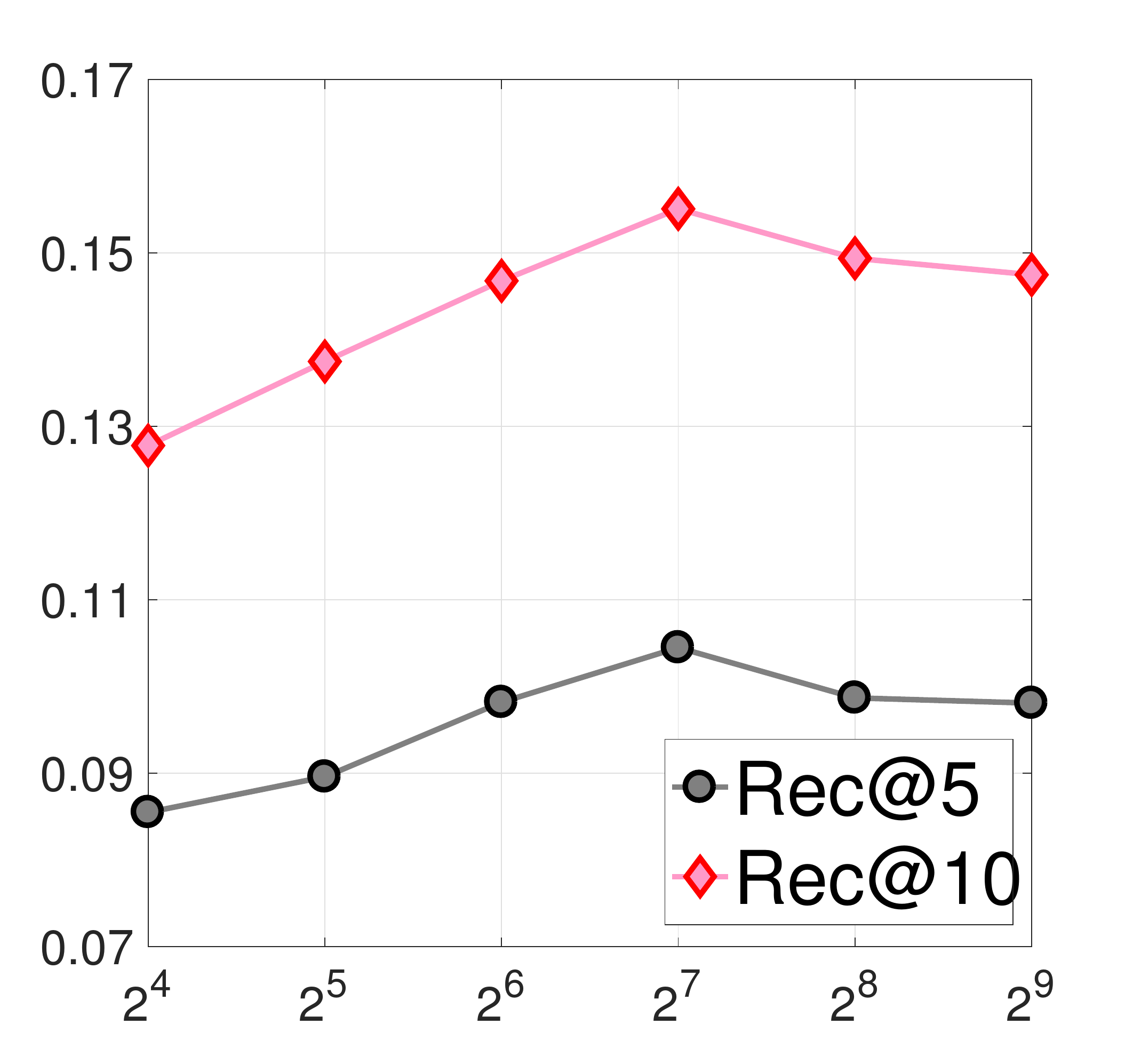}
        \vspace{-0.2in}
        \label{fig: impacts-AV-d}
        }
        \vspace{-0.2in}
    \caption{Parameter sensitivity in venue recommendation.}
     \vspace{-0.2in}
    \label{fig: impacts-AV}
\end{center}
\end{figure}

\subsubsection{Results}
The results are reported in Table \ref{tab: result-AV}. 
From the table: (a) All variants of CARL achieve better performance than baselines, showing the benefit of incorporating unstructured semantic content for learning author and venue representations; (b) The average improvements of CARL$^{mw}_{SE-HSG}$ over different baselines are significant, and range from 18.9\% to 178.7\% and 11.7\% to 275.6\% on AMiner-$\uppercase\expandafter{\romannumeral1}$ and AMiner-$\uppercase\expandafter{\romannumeral2}$; (c) The results of different variants of CARL are close due to relative small recall values. However, we find that CARL$^{rw}_{HSG-SR}$ is the worst among four, indicating both meta-path walk and semantic enhanced SkipGram help improve the performance of CARL in the venue recommendation task.

\subsubsection{Parameter Sensitivity}
Figure \ref{fig: impacts-AV} shows the impact of window size $\tau$ and feature dimension $d$ on the performance of CARL$^{mw}_{SE-HSG}$ on AMiner-$\uppercase\expandafter{\romannumeral2}$ (T = 2013). Accordingly, 
CARL$^{mw}_{SE-HSG}$ achieves the best results when $\tau$ is around 7 and $d$ is around 128 for the venue recommendation task. 


\subsection{Relevance Search: Case Study (\textbf{RQ2})} 
To answer \textbf{RQ2}, we present three case studies of relevance search on AMiner-$\uppercase\expandafter{\romannumeral2}$ (T = 2013) to show the performance differences between CARL$^{mw}_{SE-HSG}$ and baselines. The ranking of each search result is based on the cosine similarity of representations.  

\begin{table*}[!tb]
\caption{Case study of relevant author search. ``Coauthor-b'' denotes whether two authors have a collaboration relation before T and ``Similar-I'' represents whether two authors have similar research interests.}
\vspace{-0.18in}
\begin{center}
\small
\begin{tabular}{c||c|c|c|c|c|c|c|c|c}
  \toprule
\multicolumn{10}{c}{Query: Jure Leskovec (2009 SIGKDD Dissertation Award, Research Interest: Network Mining \& Social Computing)} \\
  \midrule
\multirow{2}{*}{Rank}&\multicolumn{3}{c|}{node2vec}&\multicolumn{3}{c|}{metapath2vec}& \multicolumn{3}{c}{CARL$^{mw}_{SE-HSG}$} \\
 \cmidrule{2-10}
  &Author&Coauthor-b ? &Similar-I ?&Author&Coauthor-b ? &Similar-I ?&Author &Coauthor-b ? &Similar-I ? \\
 \midrule
  \midrule
 1 &S. Kairam &{\cmark} &{\cmark} &L. Backstrom & {\cmark}&{\cmark} &J. Kleinberg & {\cmark} & {\cmark}\\
 \cmidrule{1-10}
 2 &M. Rodriguez &{\cmark} &{\cmark} &P. Nguyen &{\xmark} &{\xmark} &D. Romero&{\xmark} & {\cmark}\\
\cmidrule{1-10}
3 & D. Wang&{\cmark} & {\cmark}&S. Hanhijärvi &{\xmark} & {\xmark}&A. Dasgupta & {\cmark} & {\cmark}\\
 \cmidrule{1-10}
 4 &J. Yang &{\cmark} &{\cmark} &S. Myers &{\cmark} &{\cmark} &L. Backstrom&  {\cmark}& {\cmark}\\
\cmidrule{1-10}
 5 & A. Jaimes& {\xmark}& {\xmark}&V. Lee &{\xmark} &{\cmark} &G. Kossinets&{\xmark} & {\cmark}\\
        \midrule
  \end{tabular}
\end{center}
\vspace{-0.18in}
\label{tab: case-AA}
\end{table*}%

\begin{table*}[!tb]
\caption{Case study of relevant paper search.}
\vspace{-0.18in}
\begin{center}
\footnotesize
\begin{tabular}{c|c||c}
  \toprule
\multicolumn{3}{c}{Query: When will it happen?: relationship prediction in heterogeneous information networks (WSDM2012, Citation > 180), A: Y. Sun, J. Han,  C. Aggarwal, N. Chawla} \\
  \midrule
Model&Rank&Returned Paper \\
\midrule
\midrule
\multirow{7}{*}{node2vec}& \multirow{1}{*}{1} &Co-author relationship prediction in heterogeneous bibliographic networks (ASONAM2011), A: Y. Sun, R. Barber, M. Gupta, C. Aggarwal, J. Han\\
 \cmidrule{2-3}
 &\multirow{1}{*}{2} &A framework for classification and segmentation of massive audio data streams (KDD2007), A: C. Aggarwal\\
\cmidrule{2-3}
&\multirow{1}{*}{3} & Mining heterogeneous information networks: the next frontier (KDD2012), A: J. Han\\
 \cmidrule{2-3}
& \multirow{1}{*}{4} &Ranking-based classification of heterogeneous information networks (KDD2011), A: M. Ji, J. Han, M. Danilevsky\\
\cmidrule{2-3}
&\multirow{1}{*}{5} &Evolutionary clustering and analysis of bibliographic networks (ASONAM2011), A: M. Gupta, C. Aggarwal, J. Han, Y. Sun  \\
\midrule
\multirow{7}{*}{CARL$^{mw}_{SE-HSG}$}& \multirow{1}{*}{1} &Collective prediction of multiple types of links in heterogeneous information networks (ICDM2014), A: B. Cao, X. Kong, P. Yu\\
 \cmidrule{2-3}
 &\multirow{1}{*}{2} &Community detection in incomplete information networks (WWW2012), A: W. Lin, X. Kong, P. Yu, Q. Wu, Y. Jia, C. Li	\\
\cmidrule{2-3}
&\multirow{1}{*}{3} & Meta path-based collective classification in heterogeneous information networks (CIKM2012), A: X. Kong, P. Yu, Y. Ding, D. Wild\\
 \cmidrule{2-3}
& \multirow{1}{*}{4} &Ranking-based classification of heterogeneous information networks (KDD2011), A: M. Ji, J. Han, M. Danilevsky	 \\
\cmidrule{2-3}
&\multirow{1}{*}{5} &Fast computation of SimRank for static and dynamic information networks (EDBT2010), A: C. Li, J. Han, G. He, X. Jin, Y. Sun, Y. Yu, T. Wu	 \\
\midrule
  \end{tabular}
\end{center}
\vspace{-0.18in}
\label{tab: case-PP}
\end{table*}%

\subsubsection{Relevant Author Search}
Table \ref{tab: case-AA} lists the top-5 returned authors for query author \textbf{``Jure Leskovec''} of CARL$^{mw}_{SE-HSG}$ and two baselines, i.e., node2vec and metapath2vec, which achieves relatively better performances in the collaboration prediction task. According to this table: (a) most of returned authors of node2vec have collaboration relations with ``Jure'' before T, indicating that node2vec highly depends on structural closeness and cannot find relevant authors who are far away from ``Jure'' in the network; (b) metapath2vec returns some authors (e.g., P. Nguyen) who are different from ``Jure'' in their specific research interests, illustrating that ``APVPA'' meta-path walks (used by metapath2vec) may collect context node that are different from target node since it is common that authors bridged by the same venue study different research topics; (c) CARL$^{mw}_{SE-HSG}$ not only returns structurally close authors who have collaboration relations with ``Jure'' before T but also finds farther authors (e.g., D. Romero) who share similar research interest with ``Jure'', demonstrating CARL$^{mw}_{SE-HSG}$ captures both structural closeness and unstructured semantic relations for learning author representations. 

\subsubsection{Relevant Paper Search}
Table \ref{tab: case-PP} lists the top-5 returned papers for query paper \textbf{``When will it happen?''} of CARL$^{mw}_{SE-HSG}$ and the best baseline node2vec in paper retrieval task. From this table: (a) all returned papers of node2vec are written by at least one author in query paper, showing that node2vec only returns structurally close papers but has difficulty finding farther semantically related papers; (b) CARL$^{mw}_{SE-HSG}$ not only returns structurally close papers which have common authors with query paper but also finds semantically related papers without authorship overlapping, showing that CARL$^{mw}_{SE-HSG}$ utilizes both structural content and unstructured semantic content for learning paper representations.   

\subsubsection{Relevant Author-Venue Search}
Table \ref{tab: case-AV} lists the top-10 returned venues for query author \textbf{``Chi Wang''} of CARL$^{mw}_{SE-HSG}$ and two baselines, i.e., Deepwalk and node2vec, which have relatively better performances on the venue recommendation task. We find that: (a) Deepwalk and node2vec recommend both data mining (e.g., KDD \& ICDM) and database (e.g., SIGMOD \& PVLDB ) venues to ``Chi'' since some of his works cite database papers and some of his co-authors focus on database research, which illustrates that both Deepwalk and node2vec return structurally close venues, some of which are not the most suitable ones; (b) most of the venues in CARL$^{mw}_{SE-HSG}$'s recommendation list belong to data mining related areas, demonstrating that incorporating semantic content helps learn better representations of author and venue. 

\begin{table}[!tb]
\caption{Case study of relevant author-venue search.}
\vspace{-0.18in}
\begin{center}
\small
\begin{tabular}{c||c|c|c}
  \toprule
\multicolumn{4}{c}{Query: Chi Wang (2015 SIGKDD Dissertation Award)}\\
\multicolumn{4}{c}{Research Interest: Unstructured Data/Text Mining}\\
 \midrule
Rank&  Deepwalk& node2vec&CARL$^{mw}_{SE-HSG}$  \\
 \midrule
    \midrule
 1&KDD & KDD&  KDD\\
 2& CIKM& WSDM&CIKM \\
 3& DASFAA& SIGMOD & ICDM\\
 4&SIGMOD &PVLDB & PAKDD\\
5&ICDM &WWW  &WWW\\
 6&WSDM &ICDM& WWWC\\
 7&ICDE &EDBT &TKDE\\
 8&PVLDB &PAKDD &KAIS\\
 9&WWW &GRC &DASFAA\\
 10&EDBT& FPGA&WSDM\\
 \midrule
  \end{tabular}
\end{center}
\vspace{-0.2in}
\label{tab: case-AV}
\end{table}%

\subsection{Category Visualization of New Nodes (\textbf{RQ3})}
\begin{figure}
\begin{center}
     \subfigure[][Visualization of New Papers]{
        \centering
        \includegraphics[scale=0.4]{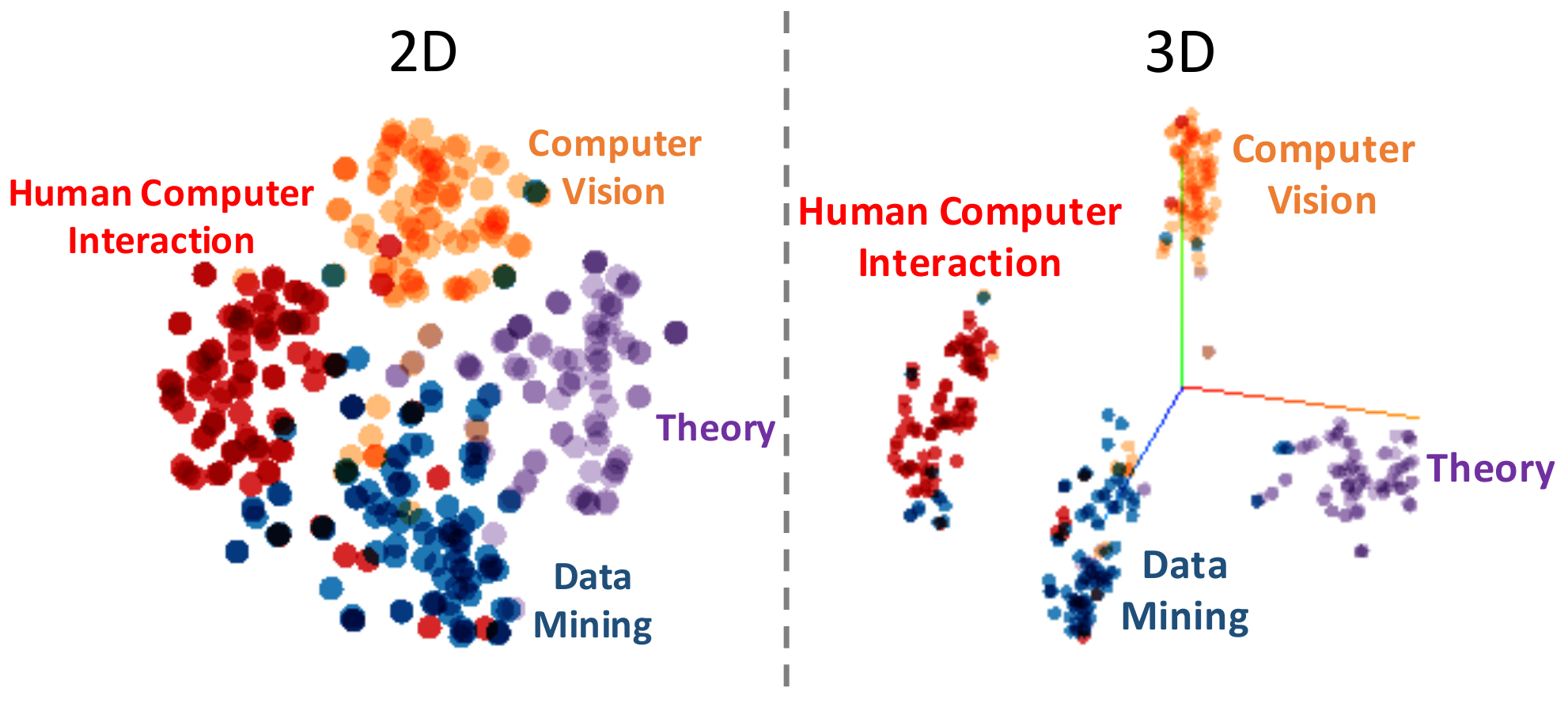}
         \vspace{-0.2in}
        \label{fig: new-p-visual}
        }
    \subfigure[][Visualization of New Authors]{
        \centering
        \includegraphics[scale=0.33]{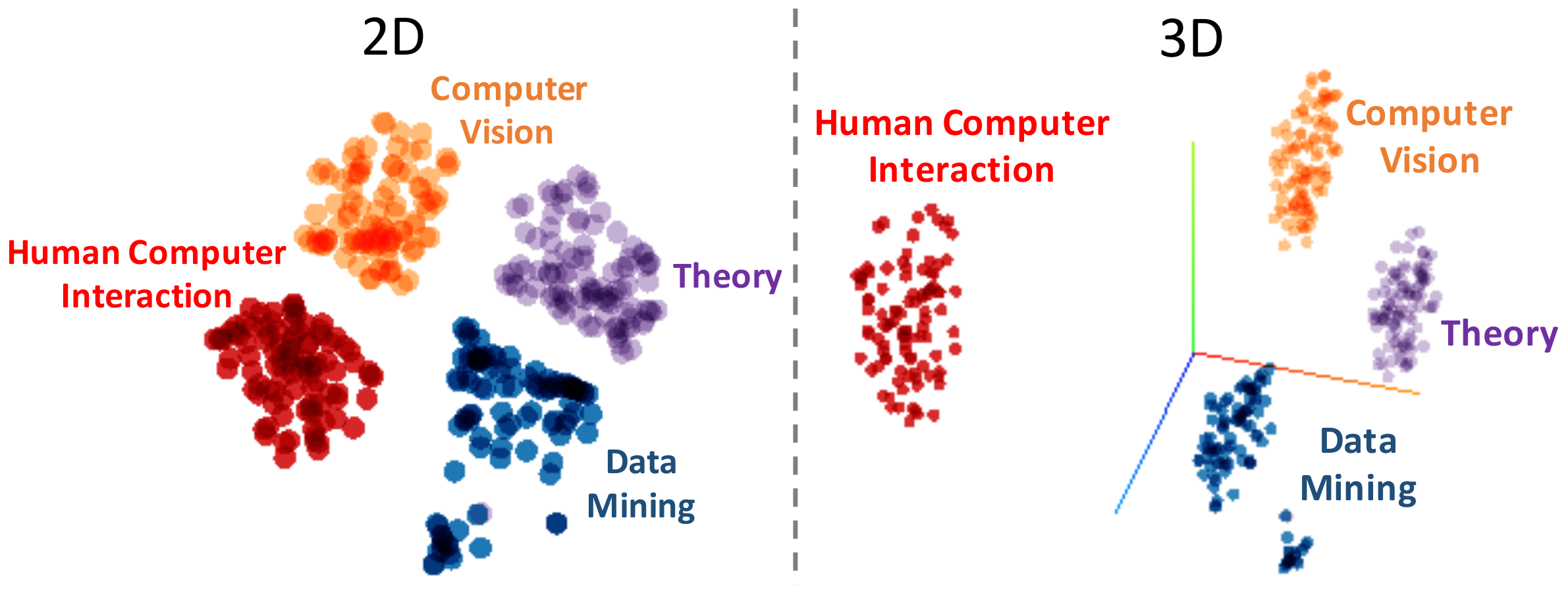}
         \vspace{-0.2in}
        \label{fig: new-a-visual}
        }
        \vspace{-0.2in}
    \caption{Representation visualizations of new paper and author nodes in four selected research categories.} 
     \vspace{-0.2in}
    \label{fig: new-visual}
\end{center}
\end{figure}
 The previous experiments and case studies demonstrate the effectiveness of CARL in learning representations of current nodes, i.e., nodes that exist in HetNet before T. 
 As described in Section 3.4, the learned semantic encoder of CARL can be directly applied to infer representation of each incoming node with semantic content. Besides, CARL's online update module can efficiently learn representation of each new node without semantic content. To answer \textbf{RQ3} and show the effectiveness of learned new node representations, we employ the Tensorflow embedding projector to visualize the new paper representations inferred by paper semantic encoder and new author representations learned by online update module in sequence. Figure \ref{fig: new-visual} shows the results of new paper and author nodes (appearing after T) of four selected research categories, i.e., Data Mining (DM), Computer Vision (CV), Human Computer Interaction (HCI) and Theory, 
 on AMiner-$\uppercase\expandafter{\romannumeral2}$ (T = 2013). Specifically, we choose three top venues\footnote{\scriptsize{DM: KDD, WSDM, ICDM. CV: CVPR, ICCV, ECCV. HCI:  CHI, CSCW, UIST. T: SODA, STOC, FOCS}} for each area. Each new paper is assigned according to venue's area and the category of each author is assigned to the area with the majority of his/her publications. We randomly sample 100 new papers/authors of each area. According to Figure \ref{fig: new-visual}: (a) The representations of new papers in the same category cluster closely and can be well discriminated from others for both 2D and 3D visualizations, indicating that the semantic encoder achieves satisfactory performance in inferring semantic representations of new papers. Notice that, papers belong to DM have few intersections with the other three since semantic content of few DM papers are quite similar to CV, HCI and Theory papers w.r.t. model, application and theoretical basis, respectively. (b) The representations of new authors in the same category are clearly discriminated from others without intersection for both 2D and 3D visualizations, which demonstrates the effectiveness of online update module in learning new author representations. 


\section{Related Work}\label{sec:related}
In the past decade, many works have been devoted to mining HetNets for different applications, such as relevance search \cite{sun2011pathsim,huang2016meta,chen2017task,zhang2018camel}, node clustering \cite{sun2009ranking,sun2012pathselclus}, personalized recommendation \cite{yu2014personalized,ren2014cluscite,yang2017bridging}.

The network representation learning has gained a lot of attention in the last few years. Some walk sampling based models \cite{perozzi2014deepwalk,grover2016node2vec,dong2017metapath2vec} have been proposed to learn vectorized node representations that can be further utilized in various tasks in network. Specifically, inspired by word2vec \cite{mikolov2013distributed} for learning distributed representations of words in text corpus, Perozzi et al. developed the innovative Deepwalk \cite{perozzi2014deepwalk} which introduces node-context concept in network (analogy to word-context) and feeds a set of random walks over network (analogy to ``sentences'') to SkipGram for learning node representations. In order to deal with neighborhood diversity, Grover \& Leskovec suggested taking biased random walks (a mixture of BFS and DFS) as the input of SkipGram.  
More recently, we argued that those models are not able to truly tackle network heterogeneity and proposed metapath2vec \cite{dong2017metapath2vec} for heterogeneous network representation learning by feeding meta-path walks to SkipGram. In addition, many other models have been proposed
\cite{tang2015line,tang2015pte,chang2015heterogeneous,ou2016asymmetric,wang2016structural,tu2018structural,wang2018graphgan,qiu2018network}, such as PTE \cite{tang2015pte} for text data embedding, HNE \cite{chang2015heterogeneous} for image-text data embedding and NetMF \cite{qiu2018network} for unifying network representation models as matrix factorization.  

Our work furthers the investigation of network representation learning by developing a content-aware representation learning model CARL for HetNets. Unlike previous models, CARL leverages both structural closeness and unstructured semantic relations to learn node representations, and contains an online update module for learning representations of new nodes.  

\section{Conclusion}\label{sec:conclusion}
In this paper, we formalize the problem of content-aware representation learning in HetNets and propose a novel model CARL to solve the problem. CARL performs joint optimization of heterogeneous SkipGram and deep semantic encoding for capturing both structural closeness and unstructured semantic relations in a HetNet. Furthermore, an online update module is designed to efficiently learn representations of new nodes. Extensive experiments demonstrate that CARL outperforms state-of-the-art baselines in various HetNet mining tasks, such as link prediction, document retrieval, node recommendation and relevance search. Besides, the online update module achieves satisfactory performance, as reflected by category visualization of new nodes. In the future, we plan to design the dynamic heterogeneous network representation learning models by using time series information of nodes and links.  
\bibliographystyle{ACM-Reference-Format}
\bibliography{reference} 

\end{document}